\documentclass{article}

\usepackage{array}
\usepackage{hyperref}
\hypersetup{colorlinks=false}
\usepackage{times}
\usepackage{t1enc}

\begin{document}

\title{Using Automated Dependency Analysis To Generate Representation Information}
\author{
Andrew N. Jackson\\
       Andrew.Jackson@bl.uk
}

\maketitle

\begin{abstract}
To preserve access to digital content, we must preserve the representation
information that captures the intended interpretation of the data.
In particular, we must be able to capture performance dependency requirements,
i.e. to identify the other resources that are required in order for the
intended interpretation to be constructed successfully. Critically,
we must identify the digital objects that are only referenced in the
source data, but are embedded in the performance, such as fonts. This paper
describes a new technique for analysing the dynamic dependencies of
digital media, focussing on analysing the process that underlies the
performance, rather than parsing and deconstructing the source data. 
This allows the results of format-specific characterisation
tools to be verified independently, and facilitates the generation of
representation information for any digital media format, even when
no suitable characterisation tool exists.
\end{abstract}

\section{Introduction}

When attempting to preserve access to digital media, keeping the bitstreams
is not sufficient - we must also preserve information on how the bits
should be interpreted. This need is widely recognised, and this data
is referred to as Representation Information (RI) by the Open Archival
Information System (OAIS) reference model \cite{CCSDS2009}. The reference
model also recognises that software can provide valuable RI, expecially
when the source code is included. However, software is not the only
dynamic dependency that must be captured in order to preserve access.
The interpretation of a digital object may inherit further information
from the technical environment as the performance proceeds, such as passwords
or licenses for encrypted resources, default colour spaces, page dimensions
or other rendering parameters and, critically, other digital objects
that the rendering requires. This last case can include linked items
that, while only referenced in the original data, are included directly
in the performance. In the context of hypertext, the term `transclusion' 
has been coined to describe this class of included resource \cite{TheodorHolmNelson2007}.

The classic example of a transcluded resource is that of fonts. Many
document formats (PDF, DOC, etc.) only reference the fonts that should
be used to render the content via a simple name (e.g. `Symbol'), and
the confusion and damage that these potentially ambiguous references
can cause has been well documented \cite{Brown2011}. Indeed, this
is precisely why the PDF/A standard \cite{ISO19005-1:2005} requires
that all fonts, even the so-called `Postscript Standard Fonts' (e.g.
Helvetica, Times, etc.), should be embedded directly in archival documents
instead of merely referenced. Similarly, beyond fonts, there are a
wide range of local or networked resources that may be transcluded,
such as media files and plug-ins displayed in web pages, documents
and presentations, or XSD Schema referenced from XML. We must be able
to identify these different kinds of transcluded resources, so that
we can either include them as explicit RI or embed them directly in the
target item (as the PDF/A standard dictates for fonts).

Traditionally, this kind of dependency analysis has been approached
using normal characterisation techniques. Software capable of parsing
a particular format of interest is written (or re-used and modified)
to extract the data that indicates which external dependencies
may be required. Clearly, creating this type of software requires
a very detailed understanding of the particular data format, and this
demands that a significant amount of effort be expended for each format
of interest. Worse still, in many cases, direct deconstruction of
the bitstream(s) is not sufficient because the intended interpretation
deliberately depends on information held only in the wider technical
environment, i.e. the reference to the external dependency is implicit
and cannot be drawn from the data. 

This paper outlines a complementary approach, developed as part of
the SCAPE project%
\footnote{\href{http://www.scape-project.eu/}{http://www.scape-project.eu/}%
}, which shifts the focus from the data held
in the digital file(s) to the process that
underlies the performance. Instead of examining the bytes, we use
the appropriate rendering software to walk-through or simulate the
required performance. During this process we trace certain operating
system operations to determine which resources are being used, and
use this to build a detailed map of the additional RI
required for the performance, including
all transcluded resources. Critically, this method does not require
a detailed understanding of file format, and so can be used to
determine the dependencies of a wide range of media without the 
significant up-front investment that developing a specialised 
characterisation tool requires.

\section{Method}

Most modern CPUs can run under at least two operating modes: `privileged'
mode and `user' mode. Code running in privileged mode has full access
to all resources and devices, whereas code running in user mode has somewhat limited
access. This architecture means only highly-trusted code has direct
access to sensitive resources, and so attempts to ensure that any
badly-written code cannot bring the whole system to a halt, or damage
data or devices by misusing them. However, code running in user space
must be able to pass requests to devices, e.g. when saving a file
to disk, and so a bridge must be built between the user and the protected
modes. It is the responsibility of the operating system kernel to manage
this divide. To this end, the kernel provides a library of system calls 
that implement the protected mode actions that the user code needs. 

Most operating systems come with software that allows these `system
calls' to be tracked and reported during execution, thus allowing
any file system request to be noted and
stored without interfering significantly with the execution process
itself%
\footnote{The tracing does slow the execution down slightly, mostly due to the
I/O overhead of writing the trace out to disk, but the process is
otherwise unaffected. %
}. The precise details required to implement this tracing approach
therefore depend only upon the platform, i.e. upon the operating system
kernel and the software available for monitoring processes running
on that kernel. 

This monitoring technique allows all file-system resources that
are `touched' during the execution of any process to be identified, and can distinguish
between files being read and files being written to. This includes
software dependencies, both directly linked
to the original software and executed by it, as well as media resources.

Of course, this means the list of files we recover includes those
needed to simply run the software as well as those specific to a particular
digital media file. Where this causes confusion, we can separate the
two cases by, for example, running the process twice, once without
the input file and once with, and comparing the results. Alternatively,
we can first load the software alone, with no document, and then start
monitoring that running process just before we ask it to load a particular
file. The resources used by that process can then be analysed from
the time the input file was loaded, as any additional resource
requirements must occur in the wake of that event.

\subsection{Debian Linux}

On Linux, we can make use of the standard system call tracer `strace',
which is a debugging tool capable of printing out a trace of all the
system calls made by another process or program%
\footnote{\href{http://sourceforge.net/projects/strace/}{http://sourceforge.net/projects/strace/}%
}. This tool can be compiled on any operating system based on a
reasonably recent Linux kernel, and is available as a standard package
on many distributions. In this work, we used Debian Linux 6.0.2 and
the Debian strace package%
\footnote{\href{http://packages.debian.org/stable/strace}{http://packages.debian.org/stable/strace}%
}. For example, monitoring a process that opens a Type 1 Postscript (PFB) 
font file creates a trace log that looks like this:
\begin{quotation}
5336 open(\textquotedbl{}/usr/share/fonts/type1/gsfonts/

\quad{}n019004l.pfb\textquotedbl{}, O\_RDONLY) = 4

5336 read(4, \textquotedbl{}\textbackslash{}200\textbackslash{}1\textbackslash{}f\textbackslash{}5\textbackslash{}0\textbackslash{}0\%!PS-

\quad{}AdobeFont-1.0: Nimbus\textquotedbl{}..., 4096) = 4096

\qquad{}\textit{\footnotesize ...more read calls...}{\footnotesize \par}

5336 read(4, \textquotedbl{}\textquotedbl{}, 4096) = 0

5336 close(4) = 0
\end{quotation}
Access to software can also be tracked, as direct dependencies like
dynamic linked libraries (e.g. `/usr/lib/libMag-\linebreak{}ickCore.so.3') appear
in the system trace in exactly the same way as any other required
resource. As well as library calls, a process may launch secondary
`child' processes, and as launching a process also requires privileged
access, these events be tracked in much the same way (via the `fork'
or `execve' system calls). The strace program can be instructed to
track these child processes, and helpfully reports a brief summary of
the command-line arguments that we passed when a new process was launched.

\subsection{Mac OS X}

On OS X (and also Solaris, FreeBSD and some others) we can use the
DTrace tool from Sun/Oracle%
\footnote{\href{http://opensolaris.org/os/community/dtrace/}{http://opensolaris.org/os/community/dtrace/}%
}. This is similar in principle to strace, but is capable of tracking
any and all function calls during execution (not just system calls
at the kernel level). DTrace is a very powerful and complex tool,
and configuring it for our purposes would be a fairly time-consuming
activity. Fortunately, DTrace comes with a tool called `dtruss',
which pre-configures DTrace to provide essentially the same monitoring capability 
as the strace tool. The OS X kernel calls have slightly different names,
the format of the log file is slightly different, and the OS X version of DTrace is not able to log the arguments passed to child processes, but these minor differences do not prevent the dependency analysis from working.

\subsection{Windows}

Windows represents the primary platform for consumption of a wide
range of digital media, but unfortunately (despite the maturity of
the operating system) it was not possible to find a utility capable
of reliably assessing file usage. The `SysInternals Suite'%
\footnote{\href{http://technet.microsoft.com/en-gb/sysinternals/bb842062}{http://technet.microsoft.com/en-gb/sysinternals/bb842062}%
} has some utilities that can identify which files a process is currently accessing
(such as Process Explorer or Handle) and similar utilities (ProcessActivityView,
OpenedFilesView) have been published by a third-party called Nirsoft%
\footnote{\href{http://www.nirsoft.net/}{http://www.nirsoft.net/}%
}. These proved difficult to invoke as automated processes, and even
when this was successful, the results proved unreliable. Each time
the process was traced, a slightly different set of files would be
reported, and files opened for only brief times did not appear at all.
Sometimes, even the source file itself did not appear in the list,
proving that important file events were being missed. This behaviour
suggests that these programs were rapidly sampling the usage of file
resources, rather than monitoring them continuously.

An alternative tool called StraceNT%
\footnote{\href{https://github.com/ipankajg/ihpublic/}{https://github.com/ipankajg/ihpublic/}%
} provides a more promising approach, as it can explicitly intercept
system calls and so is capable of performing the continuous resource
monitoring we need. However, in its current state it is difficult
to configure and, critically, only reports the name of the library 
call, not the values of the arguments.
This means that although it can be used to tell if a file was opened,
it does not log the file name and so the resources cannot be identified.
However, the tool is open source, so might provide a useful basis
for future work.

One limited alternative on Windows is to use the Cygwin UNIX-like
environment instead of using Windows tools directly. Cygwin comes with
its own strace utility, and this has functionality very similar to
Linux strace. Unfortunately, this only works for applications built
on top of the Cygwin pseudo-kernel (e.g. the Cygwin ImageMagick package).
Running Windows software from Cygwin reports nothing useful, as the
file system calls are not being handled by the Cygwin pseudo-kernel.

\section{Results}

In this initial investigation, we looked at two example files, covering two different media formats that support transcluded resources: a PDF document and a PowerPoint presentation.

\subsection{PDF Font Dependencies}
The fonts required to render the PDF test file (the
`ANSI/NISO Z39.87 - Data Dictionary - Technical Metadata for Digital
Still Images' standards document \cite{NISO2006a}) were first established
by using a commonly available tool, pdffonts%
\footnote{Part of Xpdf: \href{http://foolabs.com/xpdf/}{http://foolabs.com/xpdf/}%
}, which is designed to parse PDF files and
look for font dependencies. This indicated that the document used six fonts, 
one of which was embedded (see Table \ref{tab:Font-dependencies} for details).

The same document was rendered via three different pieces of software,
stepping through each page in turn either manually (for Adobe Reader
or Apple Preview) or automatically. The automated approach simulated
the true rendering process by rendering each page of the PDF to a
separate image via the ImageMagick%
\footnote{\href{http://www.imagemagick.org/}{http://www.imagemagick.org/}%
} conversion commmand `convert input.pdf output.jpg'. This creates
a sequence of numbered JPG images called {}`output-\#\#\#.jpg',
one for each page.

All system calls were traced during these rendering processes, and
the files that the process opened and read were collated. These lists
were then further examined to pick out all of the dependent media
files - in this case, fonts. The reconstructed font mappings are shown
in Table \ref{tab:Font-dependencies}.

\begin{table*}
\begin{tabular}{|>{\centering}p{3cm}|>{\centering}p{2cm}|>{\raggedright}p{11cm}|}
\hline 
Tool & Operating System & List of Fonts\tabularnewline
\hline 
\hline 
pdffonts 3.02 & OS X 10.7 & Arial-BoldMT. ArialMT, Arial-ItalicMT, Arial-BoldItalicMT

TimesNewRomanPSMT, BBNPHD+SymbolMT (embedded)
\tabularnewline
\hline 
Apple Preview 5.5 & OS X 10.7 & /Library/Fonts/Microsoft/...

Arial Bold.ttf, Arial.ttf, Arial Italic.ttf, Arial Bold Italic.ttf, 

Times New Roman.ttf\tabularnewline
\hline 
Adobe Reader X (10.1.0) & OS X 10.7 & /Library/Fonts/Microsoft/...

Arial Bold.ttf, Arial.ttf, Arial Italic.ttf, Arial Bold Italic.ttf\tabularnewline
\hline 
Adobe Reader 9.4.2 & Debian Linux 6.0.2 & /usr/share/fonts/truetype/ttf-dejavu/...

DejaVuSans.ttf, DejaVuSans-Bold.ttf

/opt/Adobe/Reader9/Resource/Font/ZX\_\_\_\_\_\_.PFB\tabularnewline
\hline 
ImageMagick 6.7.1 & OS X 10.7

via MacPorts & /opt/local/share/ghostscript/9.02/Resource/Font/...

NimbusSanL-Bold, NimbusSanL-Regu, NimbusSanL-ReguItal, NimbusSanL-BoldItal,
NimbusRomNo9L-Regu
\tabularnewline
\hline 
ImageMagick 6.6.0 & Debian Linux 6.0.2 & /usr/share/fonts/type1/gsfonts/...

n019004l.pfb, n019003l.pfb, n019023l.pfb, n019024l.pfb, n021003l.pfb \tabularnewline
\hline 
ImageMagick 6.4.0 & Cygwin on WinXP & /usr/share/ghostscript/fonts/...

n019004l.pfb, n019003l.pfb, n019023l.pfb, n019024l.pfb, n021003l.pfb\tabularnewline
\hline 
\end{tabular}

\caption{\label{tab:Font-dependencies}Font dependencies of a specific PDF
document, as determined via a range of tools.}
\end{table*}

The two manual renderings on OS X gave completely identical results,
with each font declaration being matched to the appropriate Microsoft
TrueType font. The manual rendering via Adobe Reader on Debian was
more complex. The process required three font files, but comparing
the `no-file' case with the `file' case showed that the first two
(DejaVuSans and DejaVuSans-Bold) were involved only in rendering the
user interface, and not the document itself. The third file, `ZX\_\_\_\_\_\_.PFB',
was supplied with the Adobe Reader package and upon inspection was
found to be a Type 1 Postscript Multiple Master font called `Adobe Sans MM', 
which contains all the variants of a typeface that Adobe Reader uses to render 
standard or missing fonts. Adobe have presumably taken this approach
in order to ensure the standard Postscript fonts are rendered consistently across platforms,
without depending on any external software packages that are beyond
their control. 

Although the precise details and naming conventions differed between
the platforms, each of the ImageMagick simulated renderings pulled
in the essentially the same set of Type 1 PostScript files, which are the open source
(GPL-compatible license) versions of the Adobe standard fonts. This
is not immediately apparent due to the different naming conventions
using on different installations, but manual inspection quickly determined
that, for example, NimbusSanL-Bold and n019004l.pfb were essentially the same
font, but from different versions of the gsfonts package. The information 
in the system trace log made it easy to determine
how ImageMagick was invoking GhostScript, and to track down the font
mapping tables that GhostScript was using to map the PDF font names
into the available fonts. 

Interestingly, as well as revealing that these apparently identical
performances depend on different versions of different files in two 
different formats (TrueType
or Type 1 Postscript fonts), the results also show that while Apple
Preview and ImageMagick indicate that Times New Roman is a required
font (in agreement with the pdffonts results) this font is not actually
brought in during the Adobe Reader rendering processes. A detailed
examination of the source document revealed that while Times New
Roman is declared as a font dependency on one page of the document,
this appears to be an artefact inherited from an older version of
the document, as none of the text displayed on the page is actually
rendered in that font.

\subsection{PowerPoint with Linked Media}

A simple PowerPoint presentation was created in Microsoft PowerPoint
for Mac 2011 (version 14.1.2), containing some text and a single image.
When placing the image, PowerPoint was instructed to only refer to
the external file, and not embed it, simulating the default behaviour
when including large media files. The rendering process was then performed
manually, looking through the presentation while tracing the system calls.
As well as picking up all the font dependencies, the fact that the image
was being loaded from an external location could also be detected
easily. 

The presentation was then closed, and the referenced image was deleted.
When re-opening the presentation, the system call trace revealed that
PowerPoint was hunting for the missing file, guessing a number of
locations based on the original absolute pathname. This approach can
therefore be used to spot missing media referenced by PowerPoint
presentations.

\section{Conclusions}

Process monitoring and system
call tracing is a valuable analysis technique, complementary
to the more usual format-oriented approach. It enables us to perform
detailed quality assurance of existing characterisation tools, using
a completely independent approach to validate the identification of
the resources required to render a digital object. Furthermore, because
the tracing process depends only on standard system functionality,
and not on the particular software in question, it can work for all
types of digital media without developing software for each format.
As the PowerPoint example shows, the only requirement for performing 
this analysis is the provision of suitable rendering software.

Before using this approach in a production setting, it will be necessary to
test it over a wider range of documents and types of transclusion, e.g. embedded XML Schema.
In particular, the monitoring should be extended to track network requests for
resources as well as local file or software calls. Although all network activity is visible via kernel system calls, the raw socket data is at such a low level that it is extremely difficult to analyse.
Fortunately, tools like netstat%
\footnote{\href{http://en.wikipedia.org/wiki/Netstat}{http://en.wikipedia.org/wiki/Netstat}%
} and WireShark%
\footnote{\href{http://www.wireshark.org/}{http://www.wireshark.org/}%
} have been designed to solve precisely this problem, and could be deployed alongside 
system call tracing to supply the necessary
intelligence on network protocols. Beyond widening the range of resources,
extending this approach to the Windows platform would be highly desirable.
The current lack of a suitable call tracing tool is quite unfortunate, and
means that this approach cannot be applied to software that only runs
on Windows. Hopefully, StraceNT can provide a way forward.

Beyond the direct resource dependencies outlined here, this approach
could be combined with knowledge of the platform package management
system in order to build an even richer model of the representation
information network a digital object requires. For example, Debian
has a rigorous package management processes, and by looking up which
packages provide the files implicated in the rendering, we can validate
not only the required binary software packages, but also determine the location of the underlying
open source software, and even the identities of the developers and other individuals
involved. This allows very rich RI to be generated
in an automated fashion. Furthermore, as the Debian package management
infrastructure also tracks the development and discontinuation of
the various software packages, this information could be leveraged
to help build a semi-automatic preservation watch system.

\section{Acknowledgments}
This work was partially supported by the SCAPE Project. The SCAPE
project is co-funded by the European Union under FP7 ICT-2009.4.1
(Grant Agreement number 270137).

\bibliographystyle{abbrv}
\bibliography{strender-local}

\end{document}